\documentclass[%
 aip,
rsi,%
 amsmath,amssymb,
 reprint,%
]{revtex4-1}

\usepackage{graphicx}% Include figure files
\usepackage{dcolumn}% Align table columns on decimal point
\usepackage{bm}% bold math
\begin{document}

\preprint{AIP/123-QED}

\title{Random magnetic anisotropy in a new compound $\mbox{Sm}_2\mbox{Ag}\mbox{Si}_3$}

\author{Baidyanath Sahu}
%\altaffiliation{Highly Correlated Matter Research Group, Department of Physics, University of Johannesburg, PO Box 524, Auckland Park 2006, South Africa}
\email{baidyanathsahu@gmail.com}
 %Lines break automatically or can be forced with \\
\author{R. Djoumessi Fobasso}%
 %\email{Second.Author@institution.edu.}
\affiliation{Highly Correlated Matter Research Group, Department of Physics, University of Johannesburg, PO Box 524, Auckland Park 2006, South Africa%\\This line break forced with \textbackslash\textbackslash
}%
\author{Andr\'{e} M. Strydom}%
 %\email{Second.Author@institution.edu.}
\affiliation{Highly Correlated Matter Research Group, Department of Physics, University of Johannesburg, PO Box 524, Auckland Park 2006, South Africa%\\This line break forced with \textbackslash\textbackslash
}%
%\author{Pranaba Kishor Muduli}
% \homepage{http://www.Second.institution.edu/~Charlie.Author.}
%\affiliation{Department of Physics, Indian Institute of Technology, Hauz Khas, New Delhi-110016, India%\\This line break forced% with \\
%}%

%\date{\today}% It is always \today, today,
             %  but any date may be explicitly specified

\begin{abstract}
We report the experimental study on the structural and magnetic properties of a new ternary intermetallic compound $\mathrm{Sm_2AgSi_3}$. The properties of the sample were investigated in detail by X-ray diffraction, dc$\textendash$magnetization, and heat capacity  measurements. The polycrystalline compound of $\mathrm{Sm_2AgSi_3}$ crystallizes in the $\mathrm{ThSi_2}$-type tetragonal structure (space group ${I_{4_1}/amd}$). The temperature$\textendash$dependent dc$\textendash$magnetization and heat capacity results demonstrate that the compound undergoes ferromagnetic behaviour with a Curie temperature ($T_C$) of 14 K. The large coercive field in hysteresis loops and the thermomagnetic irreversibility in the ferromagnetic region revealed that the compound exhibits a large magnetic anisotropy. The magnitude of the applied field and the coercivity obtained from the M-H loops corroborate with the thermomagnetic irreversibility in the magnetization data. The magnetic contribution of the heat capacity reveals a broad Schottky$\textendash$type anomaly above $T_C$, due to the presence of the crystal electric field effect of $\mathrm{Sm^{3+}}$ in $\mathrm{Sm_2AgSi_3}$.

 %
%Valid PACS numbers may be entered using the \verb+\pacs{#1}+ command.
 \end{abstract}

% \pacs{Valid PACS appear here}% PACS, the Physics and Astronomy
%                              % Classification Scheme.
 %Use showkeys class option if keyword
%                               %display desired
\maketitle

% \begin{quotation}
% The ``lead paragraph'' is encapsulated with the \LaTeX\ 
% \verb+quotation+ environment and is formatted as a single paragraph before the first section heading. 
% (The \verb+quotation+ environment reverts to its usual meaning after the first sectioning command.) 
% Note that numbered references are allowed in the lead paragraph.
% %
% The lead paragraph will only be found in an article being prepared for the journal \textit{Chaos}.
% \end{quotation}

%\section{\label{sec:level1}First-level heading:\protect\\ The line break was forced \lowercase{via} \textbackslash\textbackslash}

The ternary intermetallic compounds $\mathrm{R_2TX_3}$ (R = rare earths, T = transition metals, X = Si, Ge, Al, and In), are extensively studied for their structural and magnetic properties. Most of the compounds with $\mathrm{R_2TX_3}$ stoichiometry form in $\mathrm{ThSi_2}$ type both hexagonal and tetragonal crystal structure. The magnetic R ions occupy the Th position, while the T and X ions mutually occupy the Si positions in $\mathrm{ThSi_2}$ \cite{R2TX3, Tran, R2AgGe3}. Recently, the random magnetic anisotropy in a polycrystalline $\mathrm{ThSi_2}$ type hexagonal compound $\mathrm{Sm_2NiSi_3}$ was confirmed from the temperature dependence of zero-field cooled (ZFC) and field-cooled (FC) magnetization and magnetic entropy changes \cite{Sm2NiSi3-A, Sm2NiSi3-F}.

Samarium (Sm) based intermetallic compounds have gained considerable interest because of their interesting properties such as superconductivity, magneto$\textendash$resistance, magnetic and non magnetic behaviour \cite{SC1, SC2, MR}. Sm based ternary and binary compounds can show large magnetic hysteresis due to their large magnetic anisotropy and may be used as a permanent magnet \cite{RNi4B, Anisotropy}. A strong magneto-crystalline anisotropy may be produced by $\mathrm{Sm^{3+}}$ ion as a consequence of the crystal electric field (CEF) effect acting on the 4f$\textendash$electrons. 

In recent years, the development of magnetic materials has posed a great challenge to researchers. To the best of our knowledge, the existence of $\mathrm{Sm_2AgSi_3}$ compound was not earlier reported. In this work, we have reported the detailed synthesis process and magnetic properties of a new compound $\mathrm{Sm_2AgSi_3}$. The large random magnetic anisotropy in $\mathrm{Sm_2AgSi_3}$ is also demonstrated.

\subsection{\label{sec:level2}Experimental Set-up}
A polycrystalline sample $\mathrm{Sm_2AgSi_3}$ was prepared under ultra-high purity argon atmosphere using the standard arc-melting technique. The melted ingot was turned over and remelted five times to ensure a good homogeneity. The melted sample was wrapped in a tantalum foil and sealed in a vacuum quartz tube. The tube was kept in a furnace for annealing at 1373 K for one week and then quenched in cold water. The sample was characterized by powder X-ray diffraction (XRD) using $\mathrm{CuK_\alpha}$ radiation of a Rigaku XRD machine. The Rietveld analysis of XRD patterns was carried out using FULLPROF software \cite{rietveld, fullprof}. The temperature and field dependence of dc$\textendash$magnetization measurement was performed using a Dynacool physical properties measurement system (PPMS) made by Quantum Design, USA. Heat capacity measurements were carried out using the same equipment.

\subsection{\label{sec:level2}Results and Discussion}
\subsubsection{X-ray Diffraction}
\begin{figure}
	\centering
	\includegraphics[width =3.4 in, height =2.8 in]{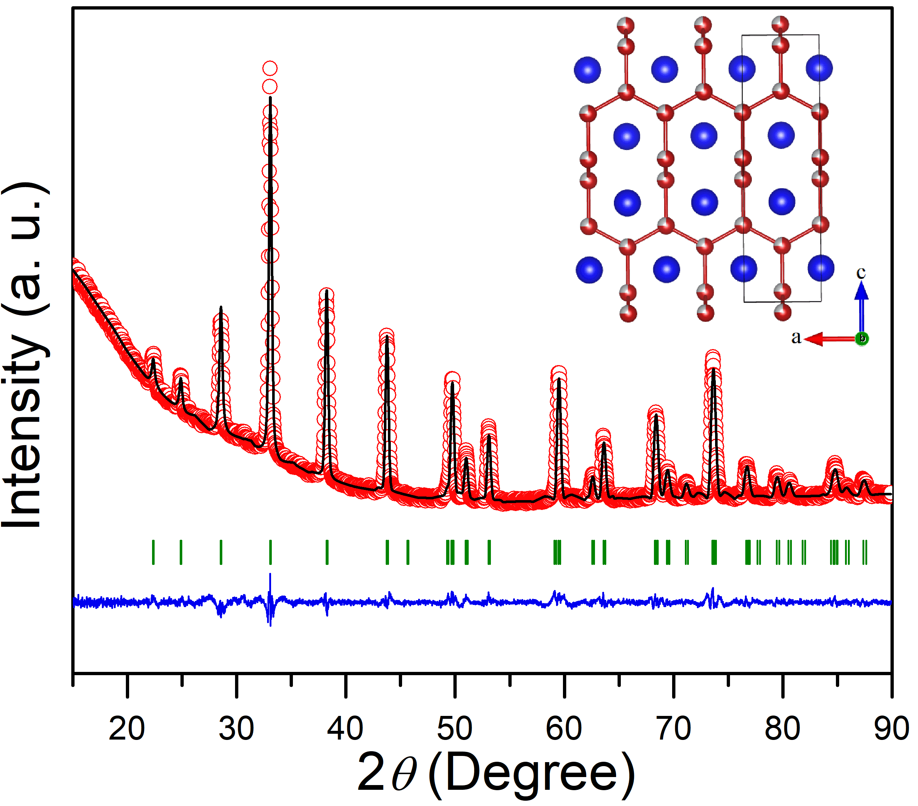}
	\caption{X-ray powder diffraction data for $\mathrm{Sm_{2}AgSi_{3}}$. Red symbols represent the experimental data and the black line represents the calculated data. The difference between experimental and calculated data is shown as a blue line. A set of vertical bars represents the Bragg peak positions of the tetragonal $\alpha$-$\mathrm{ThSi_2}$ type structure. Inset: The schematic representation of the tetragonal crystal structure of $\mathrm{Sm_2AgSi_3}$. The blue balls are for Sm and silver$\textendash$red color balls are for Ag$+$Si.}
	\label{XRD}
\end{figure}

Fig.~\ref{XRD} shows the room-temperature powder XRD pattern along with the Rietveld refinement fitting profile of polycrystalline $\mathrm{Sm_2AgSi_3}$. The Rietveld refinement result reveals that the compound form in a $\beta$$\textendash$$\mathrm{ThSi_2}$$\textendash$type of tetragonal structure with space group ${I_{{4}_{1}}/amd}$. The crystallographic parameters from the refinement are listed in Table~\ref{refinedata}. The inset of Fig.~\ref{XRD} shows the schematic diagram of the crystal structure of $\mathrm{Sm_2AgSi_3}$. The shortest distance between Sm atoms is 4.119 \AA{}, and is twice larger than the expected ionic radius of $\mathrm{Sm^{3+}}$ ($\mathrm{r_{Sm}}$ = 0.958 \AA{}). This suggests that there is a weak interaction between the rare$\textendash$earth atoms in  $\mathrm{Sm_2AgSi_3}$.

\begin{table}
\caption{\label{comparisrisionMCE} The lattice parameters, unit cell volume and the atomic coordinate positions of $\mathrm{Sm_2AgSi_3}$ obtained from the Rietveld refinement of XRD patterns. The refinement quality parameter is $\chi^2$ = 5.7.}
\begin{ruledtabular}
\begin{tabular}{ccccc}
\hspace{-1.5 in} a~=~b 	& \hspace{-0.0 in}  & \hspace{-0.5 in} & \hspace{-0.1 in}  & \hspace{-0.7 in}4.128(2) \AA    \\
\hspace{-1.5 in} c  & \hspace{-0.0 in}  & \hspace{-0.5 in} & \hspace{-0.1 in}  & \hspace{-0.7 in} 14.254(2) \AA \\
\hspace{-1.5 in} V  & \hspace{-0.0 in}  & \hspace{-0.5 in} & \hspace{-0.1 in}  & \hspace{-0.7 in} 242.890(1)  \AA$^3$ \\
\hline
\\
\hspace{0.1 in} Atomic coordinates  for $\mathrm{Sm_2AgSi_3}$ 	& \hspace{-0.2 in}  & \hspace{-0.2 in} & \hspace{-0.2 in}     & \hspace{-0.2 in} \\
\hline
\\
 \hspace{-1.5 in}Atom & \hspace{-2.5 in}Wyckoff & \hspace{-1.5 in}$x$ & \hspace{-0.7 in}$y$ & \hspace{-0.2 in}$z$ \\
 \hline
 & & & & \\
 \hspace{-1.5 in}Sm & \hspace{-2.5 in}$4a$ & \hspace{-1.5 in}0 & \hspace{-0.7 in}3/4 & \hspace{-0.2 in}1/8\\
 \hspace{-1.5 in}Ag & \hspace{-2.5 in}$8e$ & \hspace{-1.5 in}0 & \hspace{-0.7 in}1/4 & \hspace{-0.2 in}0.2895(2)\\
 \hspace{-1.5 in}Si & \hspace{-2.5 in}$8e$ & \hspace{-1.5 in}0 & \hspace{-0.7 in}1/4 & \hspace{-0.2 in}0.2895(2)\\
\end{tabular}
\end{ruledtabular}
\label{refinedata}
\end{table}

%Rows whose columns span multiple columns can be typeset using \LaTeX's
%\verb+\multicolumn{#1}{#2}{#3}+ command
%(for example, see the first row of Table~\ref{tab:table3}).%

\subsubsection{Magnetic Properties}

\begin{figure}
	\centering
	\includegraphics[width =3.4 in, height =2.8 in]{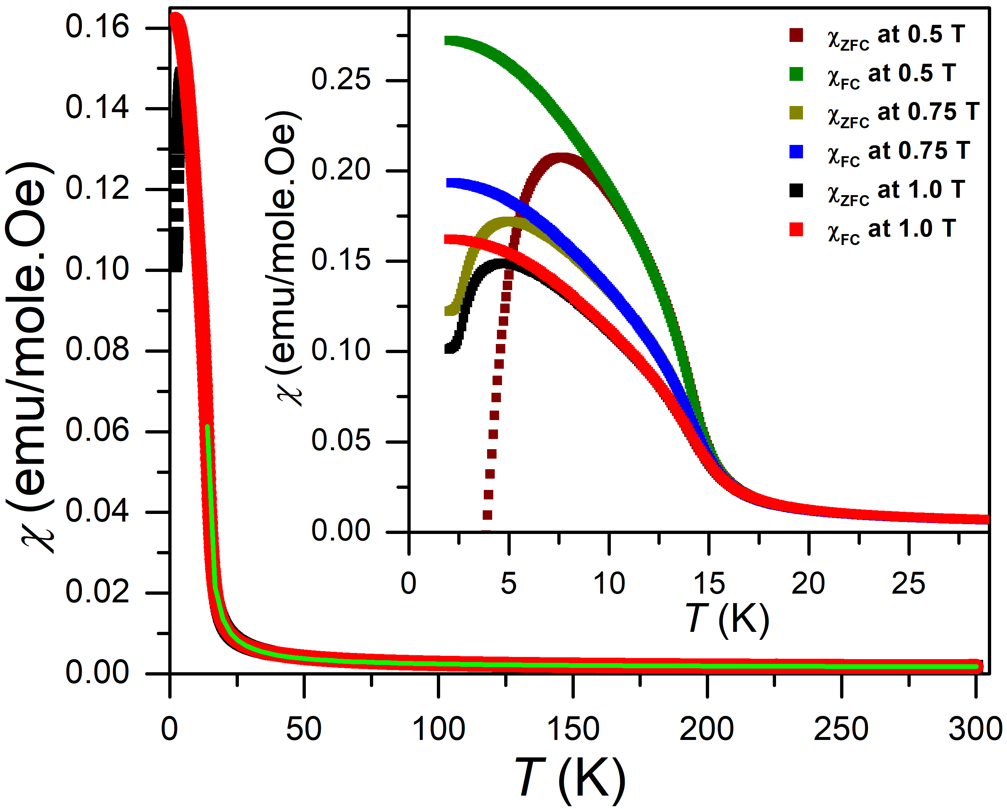}
	\caption{Temperature dependence of the field-cooled (FC) and zero$\textendash$field$\textendash$cooled (ZFC) dc$\textendash$magnetic susceptibility ($\chi(T)$) at 1 T of $\mathrm{Sm_{2}AgSi_{3}}$ along with a fit to the data (green line) of a modified Curie-Weiss law. Inset shows ZFC and FC of $\chi(T)$ at low temperature region for various applied magnetic fields.}
	\label{MT}
\end{figure}

Fig.~\ref{MT} shows the temperature dependence of ZFC and FC dc$\textendash$magnetic susceptibility ($\chi(T)$) with an applied field ($H_{app}$) of 1 T. The dc$\textendash$$\chi(T)$ shows a typical paramagnetic to ferromagnetic transition in this compound. The Curie temperature ($T_C$) was found to be 14 K from the peak of the d$\chi(T)$/d$T$ curve. The inverse dc$\textendash$magnetic susceptibility (not shown in figure) does not follow the Curie$\textendash$Weiss relation. Hence, the modified Curie$\textendash$Weiss law,  $\chi$ = [$\mathrm{C}$/($T$-$\theta_P$)]+$\chi_0$, (where C is the Curie constant, $\theta_P$ is the Weiss paramagnetic temperature and $\chi_0$ is the temperature$\textendash$independent magnetic susceptibility which includes the core-electron diamagnetism) was fitted on the $\chi(T)$ of Fig.~\ref{MT} for $T$~$\textgreater$~20 K. The  least$\textendash$squares fit (LSQ) fit to the experimental data yielded $\theta_P$ = 12 K and $\chi_0$ = 3.1 $\times$ $\mathrm{10^{-4}}$ emu/mole.Oe. The positive value of $\theta_P$ shows that the dominant interaction is ferromagnetic in $\mathrm{Sm_{2}AgSi_{3}}$. The effective magnetic moment ($\mu_{eff}$) value was calculated from the fitting parameter value of C and found to be 0.57 $\mu_B$/$\mathrm{Sm^{3+}}$. A small deviation of $\mu_{eff}$ from the theoretical value of free ion $\mathrm{Sm^{3+}}$ [$g$$\sqrt{J(J+1)}$ = 0.85 $\mu_B$] may arise from crystal field effects (CEF). The expanded region in low temperature of $\chi(T)$ for different magnetic fields shows in the inset of Fig.~\ref{MT}. As seen in the inset, there is a thermal magnetic irreversibility (TMI) shown between the data collected in the ZFC and FC protocols. The TMI point is defined as the temperature at which the $\chi_{ZFC}$ and $\chi_{FC}$ curves bifurcate from each other. It is also observed that TMI depends on the $H_{app}$. It can be believed that the large magnetic anisotropy may cause the TMI behavior in $\chi(T)$ for high $H_{app}$ = 1.0 T.

%-------------------------------------------------------------------------------------------------
%--------------------------------------------------------------------------------------------------

\begin{figure}
	\centering
	\includegraphics[scale = 0.315]{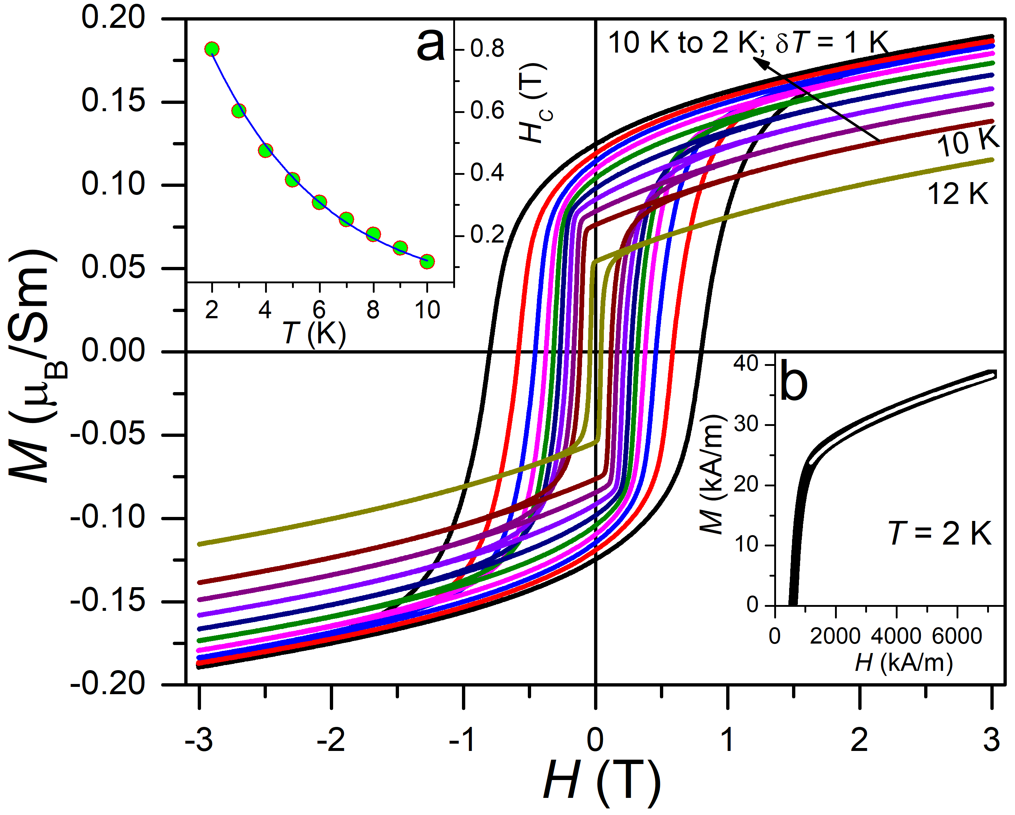}
	\caption{Hysteresis loops of $\mathrm{Sm_{2}AgSi_{3}}$ for different temperatures between 2 and 12 K. Inset (a): the temperatures dependence of the coercivity along with the fitting line explained in the text. Inset (b): the high field magnetization data with the fitted line by using Eq.~\ref{APS}.}
	\label{MH}
\end{figure}

Fig.~\ref{MH} shows the field-dependent magnetization $M(H)$ loops at different temperatures. The well$\textendash$defined hysteresis below $T_C$ is observed in the ferromagnetic ordered region of $\mathrm{Sm_{2}AgSi_{3}}$. The $M(H)$ loops do not saturate even at a high field value of 9 T at 2 K. The spontaneous magnetization was calculated from $M(H)$ at 2 K, and the obtained value is 0.25 $\mu_B$/Sm. This value is smaller by a considerable margin than the expected saturation moment value for parallel alignment of $\mathrm{Sm^{3+}}$ spin ($g$$J$ = 0.71~$\mu_B$, with $g$ = 2/7, and J = 5/2). The low magnetization value is attributed to the influence of the magnetic anisotropy in saturation. As seen in the M-H loops, a large value of coercivity ($H_C$) of 0.8 T at 2 K appears for a polycrystalline rare$\textendash$earth based compound. This large $H_C $ may result from the random magnetic anisotropy of the sample. For more details, the $M(H)$ loops were measured at different temperatures. The $H_C (T)$ and remanence magnetization was estimated from $M(H)$ loops for each temperature and found to gradually decrease with increasing temperature. The temperature variation $H_C (T)$ is plotted in inset (a) of Fig.~\ref{MH}. An exponential behavior described by the relation $H_C (T)$~=~${Aexp(-BT)}$ is observed  \cite{Pr3Ru4Al12}. The fitting is shown as a blue line on the data. The best fitting parameters obtained from the $H_C (T)$ $vs.$ $T$ curve are $A$ = 1.256 T and $B$ = 0.234 K$^{-1}$ represent the  $H_C$ at 0 K and the steep temperature respectively. The steep temperature dependence of  $H_C (T)$ in ferromagnet suggests that the observed hysteresis of $\mathrm{Sm_{2}AgSi_{3}}$ is appropriate for large magnetic anisotropy.

%--------------------------------------------------------------------------------------------------
%--------------------------------------------------------------------------------------------------

The magneto$\textendash$crystalline anisotropy constant of magnetic materials are evaluated from  the law of approach to saturation. The law of approach to saturation is therefore applied on the $M(H)$ by fitting Eq.~\ref{APS} on the high magnetic field magnetization data as  \cite{A1, A2, R2Ni7}:

\begin{eqnarray}
M = M_S \left[1-\frac{D}{H^2}\right]+\chi_0H^{1/2},
\label{APS}
\end{eqnarray}
where $M_S$ (in A/m) is the saturation magnetization. Here, $D$ is a function of magnetic anisotropy energy and $\chi_0 H^{1/2}$ denotes the term related to para-process. The parameter $D$ is expressed as:

\begin{eqnarray}
D = \frac{8}{105}\frac{K^2_1}{\mu^2_0M^2_S} \mathrm{(A/m)^2},
\label{aniso}
\end{eqnarray}
where $K_1$ is the anisotropy constant in J.m$^{-3}$ and $\mu_0$ is the free space magnetic permeability. For polycrystalline material, it is assumed that the overall possible orientations of the individual crystallites are averaged. The value of $K_1$ was estimated by using the fitting parameters $B$ and $M_{S}$ on Eq.~\ref{aniso}. The obtained $K_1 $ value is $|K_1|$ = 1.3$\times$10$^5$ J.m$^{-3}$ at 2 K. The obtained value of $|K_1|$ is comparable with 9.2$\times$10$^6$ J.m$^{-3}$ for $\mathrm{SmFe_{11}Ti}$ and 4.0$\times$10$^6$J.m$^{-3}$ for $\mathrm{Sm_3Fe_5O_{12}}$~ \cite{SmFeTi,garnet}. The observed large hysteresis loop in $M(H)$ and comparable value of $|K_1|$ revealed that a random magnetic anisotropy exists in $\mathrm{Sm_{2}AgSi_{3}}$.

The anisotropy field plays a crucial role in determining the FC magnetic susceptibilities ($\chi_{FC}$) and the ZFC magnetic susceptibilities ($\chi_{ZFC}$) at a given field value. The temperature variation of anisotropy field and  $H_{app}$ for measuring $\chi (T)$  plays a major role in determining the degree of TMI \cite{joy1}. Joy $et~al.,$ \cite{joy1,joy2} have proposed an empirical relation between $H_{app}$, $H_C$, $\chi_{FC}$ and $\chi_{ZFC}$ for a ferromagnetic system:

\begin{eqnarray}
\chi_{ZFC} = \chi_{FC}\frac{H_{app}}{H_{app}+H_C},
\label{Joy}
\end{eqnarray}   

This Eq.~\ref{Joy} indicates that the $\chi_{ZFC}$ and $\chi_{FC}$ should overlap in that temperature regime for $H_{app}$ $\gg$ $H_C$. However, the difference between the $\chi_{ZFC}$ and $\chi_{FC}$ values is large for $H_{app}$ $\ll$ $H_C$. Hence, it can be assume that the hysteresis between $\chi_{ZFC}$ and $\chi_{FC}$ arises even at high field $H_{app}$ of 1.0 T due to the magnetic anisotropy in the sample. However, the $\chi_{ZFC}$ for H =  0.5 T starting from negative values of at low temperatures is due to this anisotropic sample being cooled in a net negative trapping field. This trap magnetic field commonly exists in the PPMS superconducting solenoid magnet \cite{PPMS}.

%One can calculate $\chi_{ZFC}$ from the experimental observed value of $H_C$ and $\chi_{FC}$ by using  Eq.~\ref{Joy}. In this case, $H_C$ was measured at some selected temperature. Therefore, $\chi_{ZFC}$ was calculated at those temperature by using the corresponding $\chi_{FC}$ values. Fig.~\ref{MT2} shows the variation of calculated $\chi_{ZFC}$ and experimental measured $\chi_{ZFC}$ values with temperatures for $H_{app}$~=~0.75 T. The shape of calculated $\chi_{ZFC} (T)$ is almost same, however there is a difference in the magnitude between the calculated and experimental observed value of $\chi_{ZFC}$. This difference may arise from the large magnetic anisotropy in the compound. One more possibility for the difference between calculated and experimental value is the ZFC experiment, where the sample is not cooled under ideal conditions in zero field for vibrating$\textendash$sample magnetometer \cite{joy1}. 

%For an anisotropic sample, the presence of a small magnetic field during the cooling can result in the difference of $\chi_{ZFC}$. \textbf{However, small $H_{app}$ is not sufficient to magnetize the specimen at low temperature for a highly anisotropic ferromagnetic sample and showing resultant very small (some times negative value at low temperature) $\chi_{ZFC}$ values

\subsubsection{Heat Capacity}

\begin{figure}
	\centering
	\includegraphics[width =3.4 in, height =2.8 in]{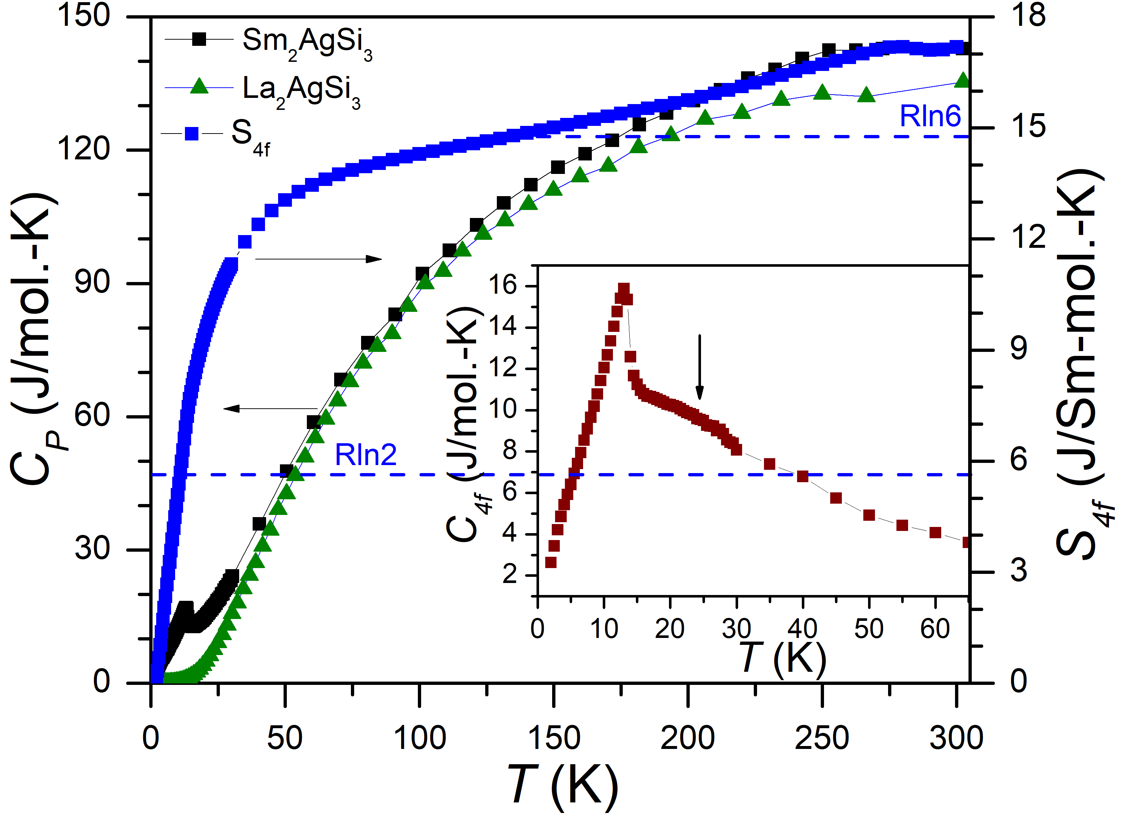}
	\caption{(Left scale) Temperature dependence of zero field heat capacity ($C_P$) of $\mathrm{La_2AgSi_3}$, and $\mathrm{Sm_2AgSi_3}$.  (Right scale) Temperature dependence of magnetic entropy of $\mathrm{Sm_2AgSi_3}$. Inset: (left with bottom scale) temperature dependence for magnetic contribution of ($C_P$) for $\mathrm{Sm_2AgSi_3}$ at low temperature region. (left with top scale) $T^{3/2}$ dependence of magnetic contribution of heat capacity for $\mathrm{Sm_2AgSi_3}$.}
	\label{CP}
\end{figure}

The temperature variation of heat capacity ($C_p(T)$) of $\mathrm{Sm_2AgSi_3}$ and of the isostructural non$\textendash$magnetic compound $\mathrm{La_2AgSi_3}$ are depicted in Fig.~\ref{CP}. A value of 147 J/(mol.$\textendash$K) is attained at 300 K, very closed to the Dulong$\textendash$Petit limit 3nR = 149.65 J/(mol.$\textendash$K) (n = 6 is the number of atoms per formula unit, R stands for the gas constant). At low temperatures, $C_p(T)$ exhibits an anomaly in $\mathrm{Sm_2AgSi_3}$ for the characteristic of a magnetic phase transition. The transition temperature of $\sim$13 K is defined from the peak of $C_p(T)$, and is consistent with the $M(T)$ data.

%The paramagnetic region of $C_P (T)$ can be derived by the standard Debye formula: 

%\begin{eqnarray}
%\nonumber
%C_P = \gamma T + 9nR\left(\frac{T}{\theta_D}\right)^3\\ \int\limits_{0}^{\theta_D/T}\frac{x^4e^x}{(e^x-1)^2}dx,
%\label{Debey}
%\end{eqnarray}
%where $\gamma$ is the Sommerfeld coefficient and $\theta_D$ is the Debye temperature. The LSQ fits of Eq.~\ref{Debey} to the experimental data yielded $\gamma$ = ---- J/(mol.$\textendash$K$^2$) and $\theta_D$ = ---- K. 

The magnetic contribution of heat capacity ($C_{4f}$) for $\mathrm{Sm_2AgSi_3}$ was estimated by subtracting the $C_P(T)$ data of $\mathrm{La_2AgSi_3}$ from that of $\mathrm{Sm_2AgSi_3}$ data and is shown in the inset of Fig.~\ref{CP}. As seen in the inset of Fig.~\ref{CP}, $C_{4f}(T)$ exhibits a broad hump above the transition temperature, indicating a Schottky type anomaly caused by crystalline electric field (CEF) present in this compound.
%The magnetically ordered region of $T^{3/2}$ dependence of $C_{4f}(T)$ (shown in inset of Fig.~\ref{CP}), may be attributed to a glassy behavior \cite{Sm2NiSi3-F}. This glassy behavior may result from either the single-ion random anisotropy or the geometrical frustration in $\mathrm{Sm_2AgSi_3}$------\textbf{we have to give some points for frustration}. 
The magnetic entropy $\rm{S_\mathrm{4f}}$ was estimated using the formula ${S_\mathrm{4f}}$~=~$\int(C_\mathrm{4f}/T)dT$. The temperature variation of ${S_\mathrm{4f}}$ is shown in the right-hand axis of Fig.~\ref{CP}. The magnetic entropy is very close to Rln2 at $\rm{T_{C}}$, indicating the presence of a doublet ground state in the magnetic system. ${S_\mathrm{4f}}$ increases above $T_C$ and displays a tendency to saturate above 50 K. ${S_\mathrm{4f}}$ saturates around a value of Rln6, indicating that the highest excited crystal field level contributes to magnetic entropy.

\subsubsection{Summary}

In this work the $\mathrm{Sm_2AgSi_3}$ compound was synthesized and found to be stoichiometric and crystallizing in the $\beta$-$\mathrm{ThSi_2}$-type tetragonal structure. The experimental results indicate that the compound undergoes a paramagnetic to ferromagnetic phase transition at $T_C$~=~14~K. The large $H_C$ confirms that the compound possesses a large large anisotropy. The presence of magnetic anisotropy plays a role in the TMI between $\chi_{ZFC}$ and $\chi_{FC}$, and are related through $H_C$. The $C_{4f}$ results indicate that the Schottky type anomaly is present due to the crystalline electric field (CEF), which may result in a large magnetic anisotropy.
The present results pave the way for future promising research of magnetic properties in new $\mathrm{R_2TX_3}$ series of compounds. 

\subsection*{\label{sec:level3}Acknowledgements}   
This work is supported by a Global Excellence and Stature (UJ-GES) fellowship, University of Johannesburg, South Africa. DFR thanks OWSD and SIDA for the fellowship towards PhD studies. AMS thanks the URC/FRC and the SA-NRF (93549) for financial support of UJ for assistance of financial support.

\subsection*{\label{ref}References}

%\nocite{*}
%\bibliography{MTJ-STO}% Produces the bibliography via BibTeX.

\begin{thebibliography}{}

	
	\bibitem{R2TX3} Z.Y. Pan, C.D. Cao, X.J. Bai, R.B. Song, J.B. Zheng, and L.B. Duan, Chinese Physics B, \textbf{22} (2013) 056102.
	\bibitem{Tran} V.H. Tran, Journal of Physics: Condensed Matter, \textbf{8} (1996) 6267.
	\bibitem{R2AgGe3}S. Sarkar, D. Mumbaraddi, P. Halappa, D. Kalsi, S. Rayaprol, and S. Peter, Journal of Solid State Chemistry, \textbf{229} (2015) 287$\textendash$295.
	\bibitem{Sm2NiSi3-A} S. Pakhira, A. K. Kundu, C. Mazumdar, and R. Ranganathan, Journal of Physics: Condensed Matter, \textbf{30} (2018) 215601.
	\bibitem{Sm2NiSi3-F} S. Pakhira, C. Mazumdar, R. Ranganathan, and S. Giri, Journal of Alloys and Compounds, \textbf{742} (2018): 391$\textendash$401.
	\bibitem{SC1}G. A. Artioli, F. Hammerath, M. C. Mozzati, P. Carretta, F. Corana, B. Mannucci, S. Margadonna, and L. Malavasi, Chemical Communications, \textbf{51} (2015) 1092$\textendash$1095.
	\bibitem{SC2} H. Fujishiro, K. Yokoyama, T. Oka, and K. Noto, Superconductor Science and Technology, \textbf{17} (2003) 51.
	\bibitem{MR} E. V. Sampathkumaran, P. L. Paulose, and R. Mallik, Physical Review B, \textbf{54} (1996) R3710.
	\bibitem{RNi4B} T. Toli\'{n}ski, A. Kowalczyk, A. Szlaferek, B. Andrzejewski, J. Kov\'{a}c, and M. Timko, Journal of Alloys and Compounds, \textbf{347} (2002) 31-35.
	\bibitem{Anisotropy} A. E. Ray, Journal of Applied Physics, \textbf{55} (1984) 2094-2096.	
	\bibitem{rietveld} H. M. Rietveld, 	Journal of Applied Crystallography, \textbf{2} (1969) 65.
	\bibitem{fullprof} Rodriguez-Carvajal J., Fullprof Suite http://www. ill. eu/sites/fullprof/ (2017).
	\bibitem{Pr3Ru4Al12} M. S. Henriques, D. I. Gorbunov, A. V. Andreev, X. Fabr\`{e}ges, A. Gukasov, M. Uhlarz, V. Pet\v{r}\'{\i}\v{c}ek, B. Ouladdiaf, and J. Wosnitza, Physical Review B, \textbf{97} (2018) 014431.	
	\bibitem{A1} S. Chikazumi,Physics of Ferromagnetism, seconded.,Oxford University Press, Oxford, 1997, p.506.
	\bibitem{A2} O. Kohmoto, Journal of Applied Physics, \textbf{53} (1982) 7486.
	\bibitem{R2Ni7} A. Bhattacharyya, S. Giri, and S. Majumdar, Journal of Magnetism and Magnetic Materials, \textbf{323} (2011) 1484$\textendash$1489.
	
	\bibitem{SmFeTi} H. T. Kim, Y. B. Kim, C. S. Kim, and H. Jin, Journal of Magnetism and Magnetic Materials, \textbf{152} (1996) 387$\textendash$390.
	
	\bibitem{garnet} K. P. Belov, A. K. Gapeev, R. Z. Levitin, A. S. Markosyan, and Yu. F. Popov, Sou. Phys.	Journal of Experimental and Theoretical Physics \textbf{41} (1975) 117.	
	\bibitem{joy1} P. A. Joy, PS Anil Kumar, and S. K. Date, Journal of Physics: Condensed Matter, \textbf{10} (1998) 11049.
	\bibitem{joy2} PS Anil Kumar, P. A. Joy, and S. K. Date, Solid State Communications, \textbf{108} (1998) 67$\textendash$70.
	\bibitem{PPMS} D. X. Li, A. D\"{o}nni, Y. Kimura, Y. Shiokawa, Y. Homma, Y.	Haga, E. Yamamoto, T. Honma, and Y. Onuki, Journal of Physics: Condensed Matter, \textbf{11} (1999) 8263.
	
	
\end{thebibliography}

\end{document}